\documentclass[preprint,onecolumn,nofootinbib]{revtex4}
\pdfoutput=1
\usepackage{float}
\usepackage[colorlinks=true,linkcolor=blue,urlcolor=blue,filecolor=black,citecolor=red,pdfstartview=FitV,pdftitle={},pdfsubject={},pdfkeywords={},pdfpagemode=None,bookmarksopen=true]{hyperref}
\usepackage{graphicx}
\usepackage{amsmath}
\usepackage{amsfonts}
\usepackage{amssymb,ulem}
\usepackage{color,xcolor}%
\usepackage{CJK}
\usepackage{subfigure}

\usepackage{dcolumn}
\newcommand{\f}{\begin{equation}}
\newcommand{\ff}{\end{equation}}
\newcommand{\fa}{\begin{eqnarray}}
\newcommand{\ffa}{\end{eqnarray}}

\begin{document}
\title{Hydrodynamic modes in holographic multiple-axion model}

\author{Ling-Zheng Xia$^{1}$}

\author{Wei-Jia Li$^{1}$}
\thanks{weijiali@dlut.edu.cn (corresponding author)}

\affiliation{$^{1}$Institute of Theoretical Physics, School of Physics, Dalian University of Technology, Dalian 116024, China.
}

\begin{abstract}
In this paper we investigate the shear viscoelasticity and the hydrodynamic modes in a holographic solid model with several sets of axions that all break the translations spontaneously on boundary. Comparing with the single-axion model, the shear modulus is enhanced at high temperatures and the shear viscosity is always suppressed in the presence of additional axions. However, the different sets of axions exhibit competitive relationship in determining the shear modulus at low temperatures. Furthermore, by calculating the black hole quasi-normal modes, it is found that adding more axions only increases the amount of diffusive modes. The number of the sound modes always remains unchanged.

\end{abstract}
\maketitle
\tableofcontents
\section{Introduction}
The concepts of symmetry and symmetry breaking play a very important role in classifying and describing different phases of matter in modern physics \cite{Nicolis:2015sra}. In condensed matter physics, especially for solid state systems, the spatial translations are ubiquitously broken either spontaneously or explicitly due to the existence of lattices, density waves, impurities or some other structures \cite{Lubensky,RevModPhys.60.1129}. 

From the conventional perspective of solid physics, the long-range order associated with the spontaneous breaking of translations endows the system with its ability to respond to mechanical deformations elastically. As the result, the propagating speeds of the associated Goldstone modes (acoustic phonons) are determined by the elastic properties of the system \cite{Landau,Lubensky}. In some cases, the translations are not broken in a unique way. A large class of typical examples are the so-called superlattices that are made from several (twisted or non-twisted) layers of two-dimensional materials each of which breaks the translations in its own manner  \cite{Smith,YCao}. Due to the more complicated structures, these systems, in general, not only modify properties of the two-dimensional materials but also enable new features and applications. However, more complicated structures also present significant challenges to the current theoretical framework. 

During the past two decades, the holographic duality (also known as AdS/CFT correspondence) has been shown to be a powerful tool in dealing with real-time dynamics of strongly correlated many-body systems \cite{Ammon_Erdmenger_2015,Zaanen_Liu_Sun_Schalm_2015,Hartnoll:2016apf,Baggioli:2019rr}. In recent years, this duality has been applied to the realm of solid state physics by breaking the translations of the boundary system spontaneously \cite{Donos:2013gda,Donos:2013wia,Jokela:2014dba,Amoretti:2016bxs,Cremonini:2016rbd,Cai:2017qdz,Alberte:2017cch,Andrade:2017cnc,Alberte:2017oqx,Amoretti:2017frz,Amoretti:2017axe,Baggioli:2022pyb}. Many progresses have been made in the studies of transport properties \cite{Ling:2014saa,Baggioli:2014roa,Baggioli:2016oqk,Gouteraux:2016wxj,Gouteraux:2018wfe,Donos:2018kkm,Amoretti:2018tzw,Baggioli:2018bfa,An:2020tkn,Andrade:2020hpu,Wu:2021mkk,Li:2022yad,Amoretti:2021fch,Liu:2022bdu,Ji:2022ovs,Ahn:2023ciq,Bai:2023use}, quantum chaos \cite{Baggioli:2016pia,Baggioli:2020ljz,Fu:2022qtz}, spectra of collective excitations \cite{Donos:2019hpp,Amoretti:2019cef,Amoretti:2019kuf,Ammon:2019wci,Baggioli:2019sio,Baggioli:2019abx,Baggioli:2019aqf,Baggioli:2019jcm,Amoretti:2020ica,Baggioli:2020edn,Baggioli:2020nay,Wang:2021jfu,Ahn:2022azl,Baggioli:2022uqb,Yang:2023vxz}, elastic response \cite{Donos:2011bh,Andrade:2019zey,Baggioli:2019elg,Baggioli:2019mck,Pan:2021cux,Baggioli:2020qdg,Baggioli:2023dfj} as well as phase transitions \cite{Baggioli:2022aft,Yang:2023dvk}. Among the models involved in these works, the most economic one is the so-called holographic axion model. It is constructed by the Einstein gravity coupled with several massless scalars whose profiles are taken to be linear in the spatial coordinates. This allows translations to be broken, while the background geometry still remains homogeneous. For a detailed review of this model, one refers to \cite{Baggioli:2021xuv}.

To enrich the structure of the boundary system, we generalize the holographic axion model in this work by introducing more axions in the bulk all of which break the translations of the dual systems spontaneously.  We aim to investigate the impacts of the additional axions on the viscoelasticity as well as the spectra of collective excitations on boundary. 

The paper is organized as follows: In section \ref{section2}, we introduce the generalized holographic model with multiple sets of axions and obtain the black hole solution. In section \ref{section3}, the shear modulus and shear viscosity of the boundary system are calculated in the framework of linear response. In section \ref{section4},the dispersion relations of collective modes are achieved by computing the low-lying quasi-normal modes (QNMs) of the black hole. Finally, we conclude in section \ref{section5}.

\section{Holographic action and black hole solution}\label{section2}
In this paper, we generalize the holographic axion model in \cite{Baggioli:2021xuv} by including multiple sets of massless scalars. And the action is given by 
\begin{equation}
    S=\int d^4x\sqrt{-g}\bigg[\frac{R}{2}-\Lambda-\mathcal{W}(X_{1},X_{2},\cdots,X_{\mathcal{N}})\bigg],
\end{equation}
where $R$ is the Ricci scalar, $\Lambda$ is the cosmological constant which we will fix $\Lambda=-3$ such that the AdS radius is a unit, $\mathcal{W}$ is a general function of $X_{a}\equiv \dfrac{1}{2}\sum_I\partial_\mu\Phi^I_{a}\partial^\mu\Phi^I_{a}$, where $I=x,y$ and $a=1,2,\cdots,{\mathcal{N}}$. Since the scalars have no mass, the theory is invariant under the global internal shift $\Phi^I_a \to \Phi^I_a+c^I_a$.

Varying the metric as well as the scalars, we get equations of motion (EOMs) as follows,
\begin{equation}
    G_{\mu\nu}-3g_{\mu\nu}+\mathcal{W}g_{\mu\nu}-\sum_{a=1}^\mathcal{N}\sum _{I=x,y}\mathcal{W}_{X_a}\partial_\mu\Phi^I_a\partial_\nu\Phi^I_a=0,
\end{equation}
\begin{equation}
    \partial_\mu(\sqrt{-g}\mathcal{W}_{X_a}\partial^\mu\Phi^I_a)=0,
\end{equation}
where $G_{\mu\nu}$ is the Einstein tensor and  $\mathcal{W}_{X_a}\equiv\partial\mathcal{W}/\partial X_a$.

It has been shown  in previous studies of holographic axion models that if the bulk profile of the scalars is imposed as
\begin{equation}\label{bgcon}
    \bar{\Phi}^I_a=\delta_i^Ix^i,
\end{equation}
the translational symmetry of the boundary system is broken, while the homogeneity as well as the isotropy are retained. Then, in the Eddington–Finkelstein (EF) coordinates, the ansatz on the black hole solution can be taken as
\begin{equation}
    ds^2=\dfrac{1}{u^2}[-f(u)dt^2-2dtdu+dx^2+dy^2],
\end{equation}
where $u\in[0,u_h]$ is the radial coordinate with $u=0$ denoting the AdS boundary and the black hole horizon is defined by $f(u=u_h)=0$. From the Einstein equation, the emblackening factor is given by
\begin{equation}
     f(u)=u^3\int_u^{u_h}\frac{3-\mathcal{W}(\bar{X}_{1},\bar{X}_{2},\cdots,\bar{X}_{\mathcal{N}})}{z^4} \mathrm{d} z,
\end{equation}
with $\bar{X_a}=z^2$ representing their background value. The entropy density can be calculated as  $s=2\pi/u_h^2$ and the Hawking temperature is given by
\begin{equation}
    T=-\frac{f'(u_h)}{4\pi}=\frac{3-\mathcal{W}(X_1^h,X_2^h,\cdots,X_\mathcal{N}^h)}{4\pi u_h},\ \text{with}\ X_a^h=u_h^2.
\end{equation}

In the following, we will calculate the excitations around the background. To do so, we shall first consider the model with two sets of scalars (we will call it double-axion model in this paper), 
\begin{equation}
    \mathcal{W}(X,Y)=m^2V(X)+n^2W(Y)+l^2Z(X,Y),
\end{equation}
where $V$, $W$ and $Z$ are general functions. One can easily generalize it into be extended to the case with more axons. Here, we have denoted $ X\equiv X_1\equiv \dfrac{1}{2}\partial_\mu\phi^I\partial^\mu\phi^I$ and $Y\equiv X_2 \equiv \dfrac{1}{2}\partial_\mu\chi^I\partial^\mu\chi^I$, $m$, $n$, $l$ are coupling parameters and the $Z(X,Y)$ term represents the interactions between the two sets of scalars.\footnote{We find that the interaction term does not change the behavior of the system qualitatively. For this reason, we will focus only on the case with $l=0$ in the rest of the main text. One can read the appendix \ref{App-d} for more details about the case with a non-zero $l$.} To perform explicit computation, we impose the following forms for $V$, $Y$ and $Z$,
\begin{equation}
    V(X)=X^M,W(Y)=Y^N, Z(X,Y)=X^PY^Q. \label{power}
\end{equation}
In the standard quantization, to break the translations on boundary in the spontaneous manner, we should further require that \cite{Baggioli:2021xuv}
\begin{equation}
    M,\  N,\  P+Q>\frac{5}{2}.
\end{equation}
In following, we will investigate the shear viscoelasticity as well as the hydrodynamic excitations by holography in such a specific model.

\section{Shear viscoelasticity}\label{section3}
In the linear response theory, the shear viscoelasticity can be extracted from the retarded Green's function of the stress tensor which at low frequency and zero momentum can be expressed as
\begin{equation}
    \mathcal{G}_{T_{xy}T_{xy}}^{(R)}(\omega,k=0)=G-i\omega\eta+\cdots,
\end{equation}
where $G$ is the shear elastic modulus and $\eta$ is the shear viscosity. 

In holography, this retarded function can be calculated by considering the metric fluctuation $\delta g_{xy}=\dfrac{1}{u^2}h_{xy}e^{-i\omega t}$ and solving the linearized Einstein equation,
\begin{equation}
          -h_{xy}  \left(2 l^2 u Z_Y+2 l^2 u Z_X+2 m^2 u V_X+2 n^2 u W_Y+2 i \omega \right)+h_{xy}'  \left(u f' -2 f +2 i u \omega \right)+u f  h_{xy}''=0.
\end{equation}
Near the AdS boundary ($u\rightarrow 0$), its asymptotic behavior is
\begin{equation}\label{hUV}
        h_{xy}=h_{xy}^{(0)}\,(1+\ldots)+h_{xy}^{(3)}\,u^3\,(1+\ldots).
\end{equation}
According to the holographic dictionary, the retarded Green's function can be read as
\begin{equation}\label{Tgreen}
\mathcal{G}_{T_{xy}T_{xy}}^{(R)}=\frac{3}{2}\,\frac{h_{xy}^{(3)}}{h_{xy}^{(0)}}.
\end{equation}
Then, the shear modulus and shear viscosity can be achieved respectively from
 \begin{equation}
G=\lim_{\omega\to0}\text{Re}\left[\mathcal{G}_{T_{xy}T_{xy}}^R(\omega,k=0)\right],
 \end{equation}
and
\begin{equation}
    \eta=-\lim_{\omega\to0}\frac1\omega \text{Im}\left[\mathcal{G}_{T_{xy}T_{xy}}^R(\omega,k=0)\right].
\end{equation}

\begin{figure}[htbp]
    \centering
    \includegraphics[width=0.48\linewidth]{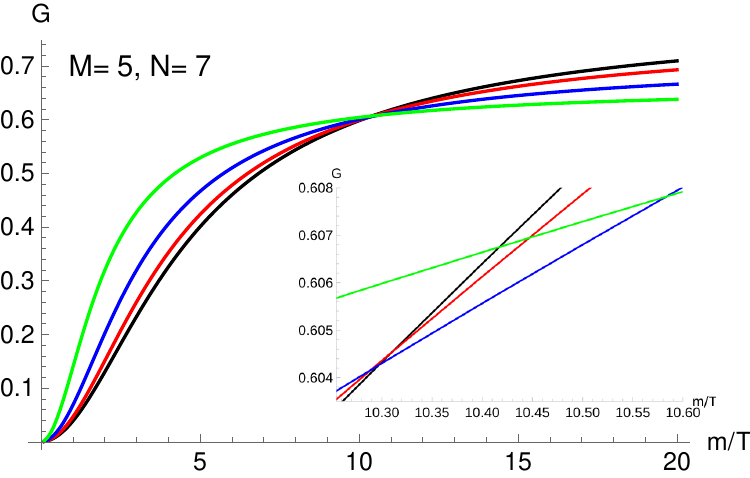}
    \includegraphics[width=0.48\linewidth]{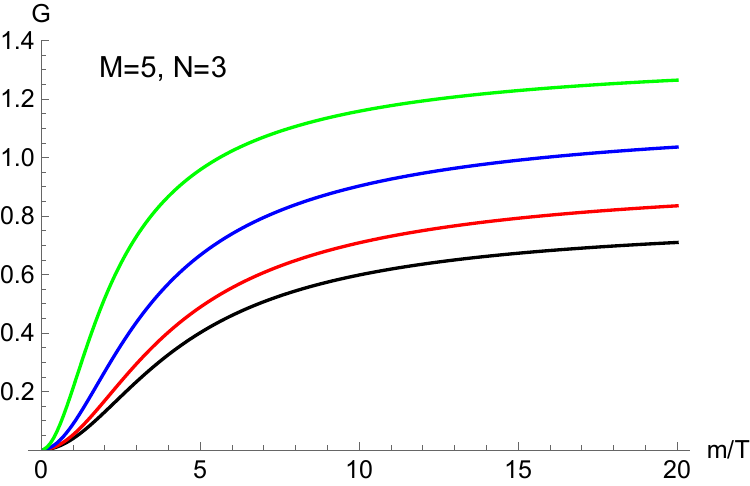}
    \caption{The shear modulus $G$ as the function of $m/T$ for $n/m =\{0,1/2,1,2\}$ (from black to green). $\textbf{Left}$: $M=5$ and $N=7$. The inserted panel shows that intersection points are not overlap. $\textbf{Right}$: $M=5$ and $N=3$.}
    \label{fig:G}
\end{figure}

In FIG. \ref{fig:G}, we plot $G$ as a function of $m/T$ for different values of $n/m$. In all cases, the shear modulus increases monotonically as the increase of $m/T$. Comparing with the single-axion model, the rigidity of the system is always enhanced by the additional scalars, $\chi^I$, at high temperatures. This can be easily explained since we have that
\begin{equation}
G\approx \int^{u_h}_0\frac{m^2V_X(\bar{X})+n^2W_Y(\bar{Y})}{z^2}\,dz,\quad \text{for}\quad m,n\ll T.
\end{equation}
However, the situation becomes different at low temperatures. In this region, for $M<N$, we find that $G$ decreases when the impact of $\chi^I$ is enhanced by increasing $n/m$. Then, at some intermediate temperature, there exists an intersection with the $G$-curve of the single-axion model. While, for $M>N$, $G$ always increases as the increase of $n/m$. It looks as if that these two sets of axions exhibit a competition instead of producing an additive effect on $G$ at low temperatures. For instance, when $m\gg n$, the $\phi^I$ fields play the dominant pole. If $m\ll n$, the $\chi^I$ fields become the most important. And as was shown in the single-axion model that larger $M$ or $N$ leads to a smaller shear modulus at low temperatures  \cite{Alberte:2017oqx}. This explains why the shear modulus in the left panel of FIG.\ref{fig:G} becomes smaller than the single axion model with $\mathcal{W}=m^2 X^5$ at low temperatures. We have verified this point in the appendix \ref{App-G}. Therefore, the result for $M<N$ and that for $M>N$ are basically the same which is consistent with the fact that the model enjoys an interchange symmetry of $X,M,m \leftrightarrow Y,N,n$. 

\begin{figure}[htbp]
    \centering
    \includegraphics[width=0.48\linewidth]{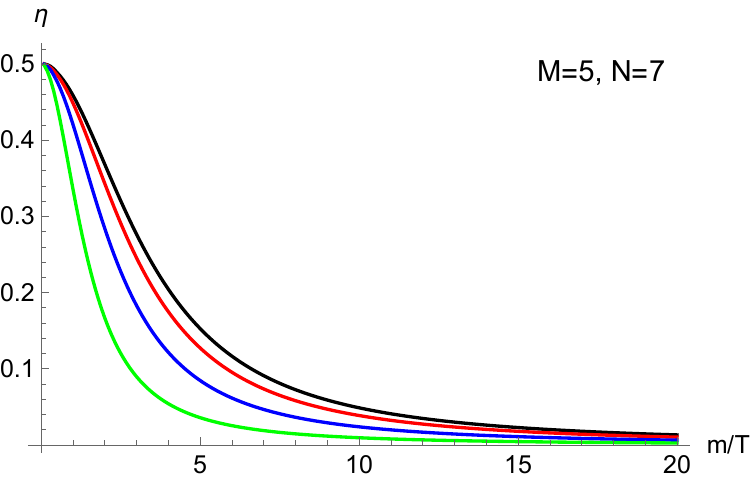}
    \includegraphics[width=0.48\linewidth]{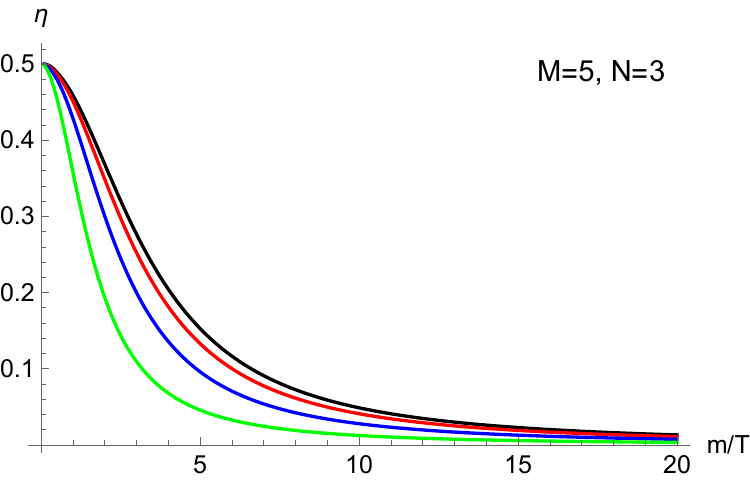}
    \caption{The shear viscosity $\eta$ as the function of $m/T$ for $n/m =\{0,1/2,1,2\}$ (from black to green). $\textbf{Left}$: $M=5$ and $N=7$. $\textbf{Right}$: $M=5$ and $N=3$.}
    \label{fig:eta}
\end{figure}

On the other hand, the shear viscosity $\eta$ as a function of $m/T$, has been shown in FIG. \ref{fig:eta}. It is obvious to see that $\eta$ monotonously decreases as the increase of $m/T$ and the additional scalars $\chi^I$ always suppress its value. 

\section{Dispersion relations of hydrodynamic modes}\label{section4}
Now, we concentrate on the dispersion relations of the hydrodynamic modes in the boundary system which can be extracted from the low-lying QNMs of black holes. Then, we need to consider perturbations of the metric and scalar fields upon the background,
\begin{equation*}
    g_{\mu\nu}=\bar{g}_{\mu\nu}+\delta g_{\mu\nu},\quad
\phi^I=\bar{\phi}^I+\delta\phi^I,\quad
    \chi^I=\bar{\chi}^I+\delta\chi^I,
\end{equation*}
where the fluctuating fields can be expanded as
\begin{equation}
\begin{aligned}
    \delta g_{\mu\nu}=\int \frac{d^3k}{(2\pi)^3}\, \dfrac{h_{\mu\nu}(\omega,k_i,u)}{u^2}\,e^{-i\omega t+ik_ix^i},\\
    \delta\phi^I=\int \frac{d^3k}{(2\pi)^3}\,  \delta^I_j\,\phi_j(\omega,k_i,u)\,e^{-i\omega t+ik_ix^i},\\
    \delta\chi^I=\int \frac{d^3k}{(2\pi)^3}\,  \delta^I_j\,\chi_j(\omega,k_i,u)e^{-i\omega t+ik_ix^i}.
\end{aligned}
\end{equation}
Since the system possesses a rotational symmetry on the $x-y$ plane, without loss of generality, one can set the wave vector $k$ to be along the $y-$direction. Then, the complete set of the fluctuating modes can be divided two sectors by the parity in the $x-$direction:
\begin{equation}
\begin{aligned}
&\text{Transverse modes (parity odd):}\quad \{\delta g_{xu},\delta g_{tx},\delta g_{xy},\delta\phi^x,\delta\chi^x\},\\
&\text{Longitudinal modes (parity even):}\quad \{\delta g_{uu},\delta g_{tu},\delta g_{yu},\delta g_{tt},\delta g_{ty},\delta g_{xx},\delta g_{yy},\delta\phi^y,\delta\chi^y\}.
\end{aligned}
\end{equation} 
Solving the linearized equations for the two sectors by imposing the infalling boundary conditions at the horizon and the normalizable (sourceless) condition on the boundary, we achieve the QNM spectra of the black hole for finite $\omega$ and finite $k$ which correspond to the poles of retarded correlators for the boundary system. And the low-lying modes give us dispersion relations of the hydrodynamic modes.

\subsection{Transverse channel}
For simplicity, we adopt the radial gauge which fixes that $\delta g_{\mu u}=0$ in the numeric computation. The equations of the transverse modes have been shown in Appendix \ref{App-a}. As the result, we find that there are a pair of sounds and a purely dissipative mode in the transverse channel whose dispersion relations at small momenta are given by
\begin{equation}
    \begin{cases}
    \text{sound mode}:&\omega=\pm v_Tk-i\dfrac{\Gamma_T}{2}k^2+\cdots,\\
    \text{diffusive mode}:&\omega=-iD_T k^2+\cdots,\\
    \end{cases}
\end{equation}
where $v_T$ is the propagating speed of the transverse sounds, $\Gamma_T$ the sound attenuation and $D_T$ is the diffusivity of the dissipative mode. 

\begin{figure}[htbp]
\centering
\includegraphics[width=0.48\linewidth]{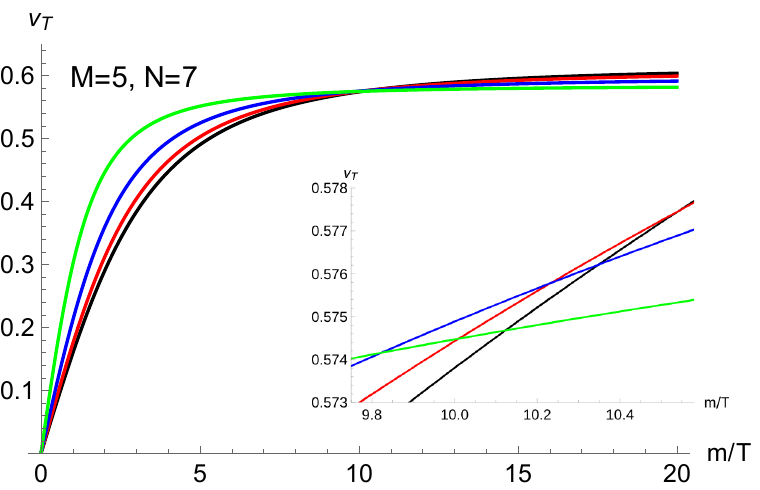}
\includegraphics[width=0.48\linewidth]{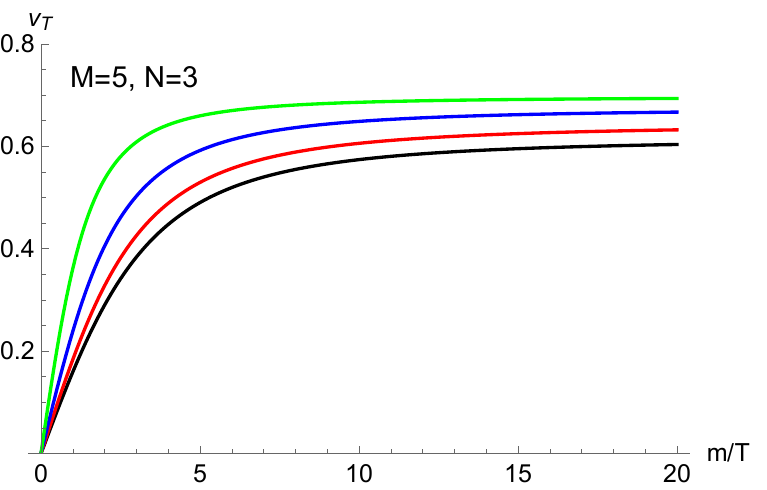}
\caption{The transverse sound velocity $v_T$ as the function of $m/T$ for $n/m =\{0,1/2,1,2\}$ (black, red, blue, green). $\mathbf{Left}$: $M=5,N=7$.$\mathbf{Right}$: $M=5,N=3$.}
\label{fig:v_T} 
\end{figure}

\begin{figure}[htbp]
\centering
\includegraphics[width=0.48\linewidth]{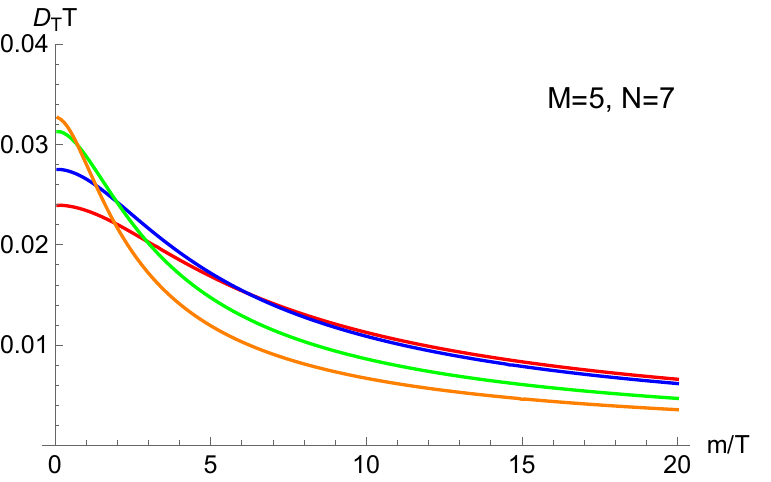}
\includegraphics[width=0.48\linewidth]{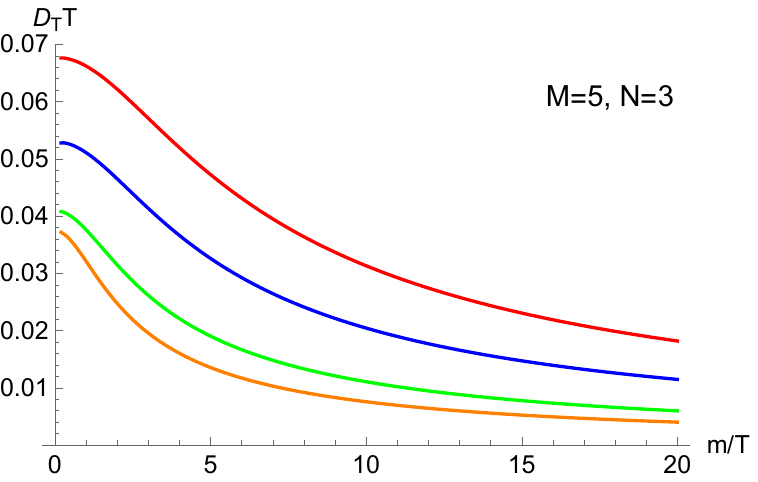}
\caption{\label{fig:Dif_T} The transverse diffusivity $D_T$ as the function of $m/T$ for $n/m =\{1/2,1,2,3\}$ (red, blue, green, orange). $\textbf{Left}$: $M=5,N=7$.  $\textbf{Right}$: $M=5,N=3$.}
\end{figure}

\begin{figure}[htbp]
    \centering
    \includegraphics[width=0.60\linewidth]{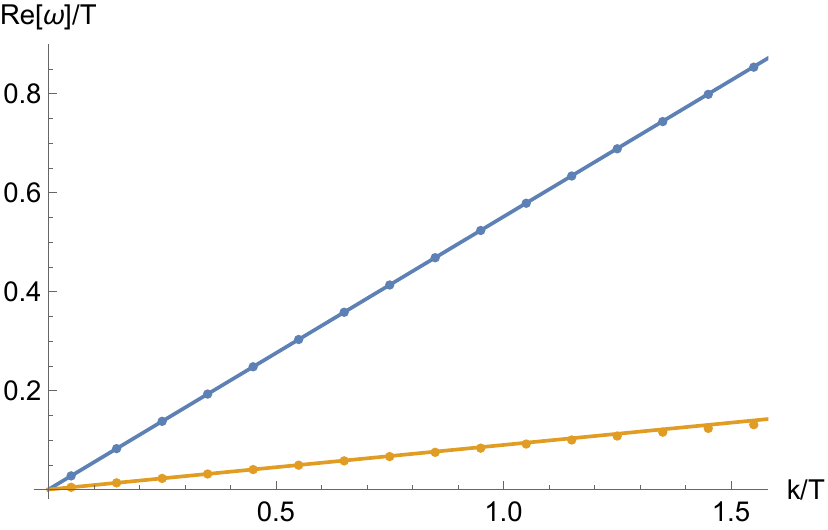}
    \caption{Comparison of the transverse sound speed extracted from the QNMs (dots) and from the analytic relation (\ref{soundmodulus}) (lines).  Blue: $n/m=1.0$ and $m/T=2.0$. Yellow: $n/m=0.5$ and $m/T=0.5$. We set $M=5$ and $N=7$.}
    \label{fig:compare}
\end{figure}

From FIG.\ref{fig:v_T}-FIG.\ref{fig:Dif_T}, it is obvious to see that, in all cases, the sound speed grows monotonously as the increase of $m/T$, while the dissipative coefficients  $\Gamma_T$ and $D_T$ behave in the opposite way. Furthermore, it has been verified in FIG.\ref{fig:compare} that the sound speed $v_T$, the shear modulus $G$ and the momentum susceptibility $\chi_{\pi\pi}$ satisfy the following relation,
\begin{equation}
    v_T=\sqrt{\frac{G}{\chi_{\pi\pi}}},\label{soundmodulus}
\end{equation}
with $\chi_{\pi\pi}=\epsilon+\mathcal{P}=\frac{3}{2}\epsilon$, where $\epsilon$ denotes the energy density which can be calculated by
\begin{equation}
    \epsilon=\frac{1}{u_h^3}\bigg(1+m^2\frac{u_h^{2M}}{2M-3}+n^2\frac{u_h^{2M}}{2N-3}\bigg),
\end{equation}
and $\mathcal{P}$ is the pressure. Note that (\ref{soundmodulus}) matches the prediction of the elasticity theory of a relativistic conformal solid, and the sounds hence can be understood as the transverse phonons.\footnote{More preciously, the transverse sounds in a solid at finite temperatures come from a combination of the transverse momentum and the Goldstone fields associated with the broken translations.} Again, we find intersections of the speed curves for $M<N$ in the left plot of FIG.\ref{fig:v_T}, which is similar to the situation of $G$ discussed in the previous section. 

The nature of the extra diffusive mode in the transverse channel is mysterious. It is irrelevant to the conserved momentum which is already coupled to the transverse Goldstone fields and gives rise to the sound pair. Its origin should be attributed to the broken global shift symmetries due to the scalar profiles (\ref{bgcon}) in the bulk \cite{Donos:2019txg}. In all cases, the dimensionless quantity $D_T T$ decreases monotonously when lowering the temperature. Moreover, we have also checked if it obeys the lower bounds on diffusion, i.e. $D \ge v_B^2\,\tau_L$ with the butterfly velocity $v_B$ and the Lyapunov time $\tau_L$,  which was firstly proposed in \cite{Blake:2016jnn}. In our model, it is found that $D_T \ll v_B^2\,\tau_L$. Therefore, the bound is strongly violated. We present this result in Appendix \ref{App-e}. Note that violation of the diffusion bound was also found in a holographic axion model where the shear mode is diffusive when the $G$ is vanishing \cite{Baggioli:2019abx}.

\begin{figure}[htbp]
\centering
\includegraphics[width=0.48\linewidth]{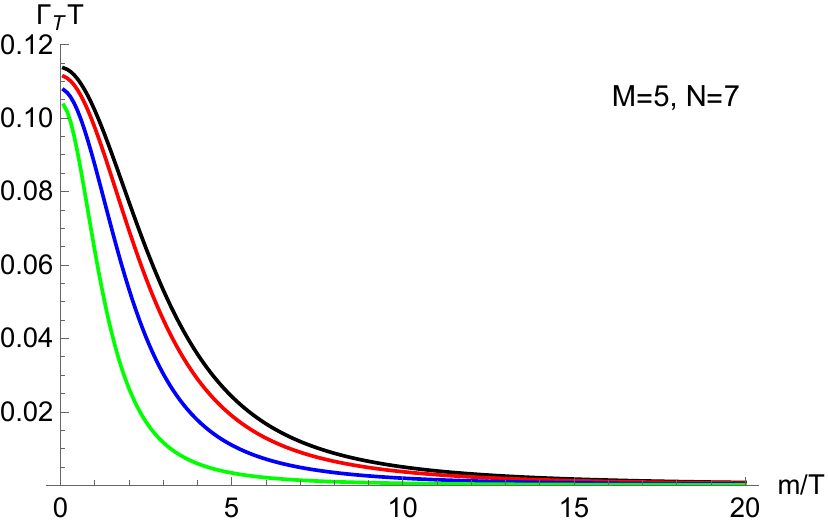}
\includegraphics[width=0.48\linewidth]{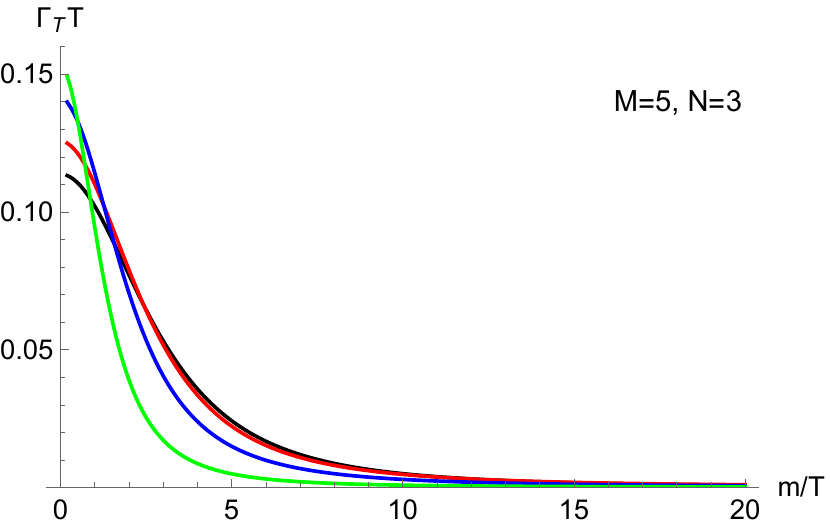}
\caption{\label{fig:D_T} The transverse sound attenuation $ \Gamma_T$ as the function of $m/T$ for $n/m =\{0,1/2,1,2\}$ (black, red, blue, green). $\textbf{Left}$: $M=5,N=7$.  $\textbf{Right}$: $M=5,N=3$.}
\end{figure}


\subsection{Longitudinal channel}

Now, we turn to study the hydrodynamic modes in the longitudinal channel. The linearized equations of the longitudinal modes have also been shown in Appendix \ref{App-a}. The dispersion relations of the hydrodynamic modes at small momenta are given by
\begin{equation}
    \begin{cases}
    \text{sound mode}:&\omega=\pm v_Lk-i\dfrac{\Gamma_L}{2}k^2+\cdots,\\
    \text{diffusive mode ``1"}:&\omega=-iD_{\phi} k^2+\cdots, \\
    \text{diffusive mode ``2"}:&\omega=-iD_{\chi} k^2+\cdots,\\
    \end{cases}
\end{equation}
where there is again an extra diffusive mode comparing with the case of single-axion model whose transport coefficient is denoted as $D_{\chi}$.

\begin{figure}[htbp]
    \centering
    \includegraphics[width=0.48\linewidth]{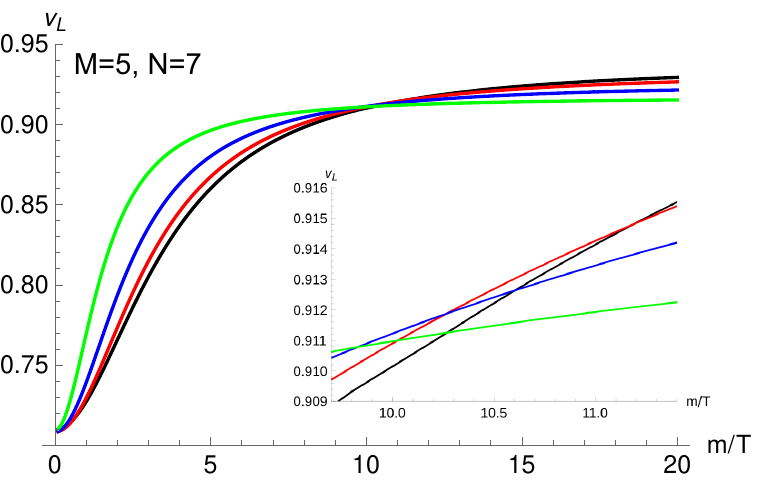}
    \includegraphics[width=0.48\linewidth]{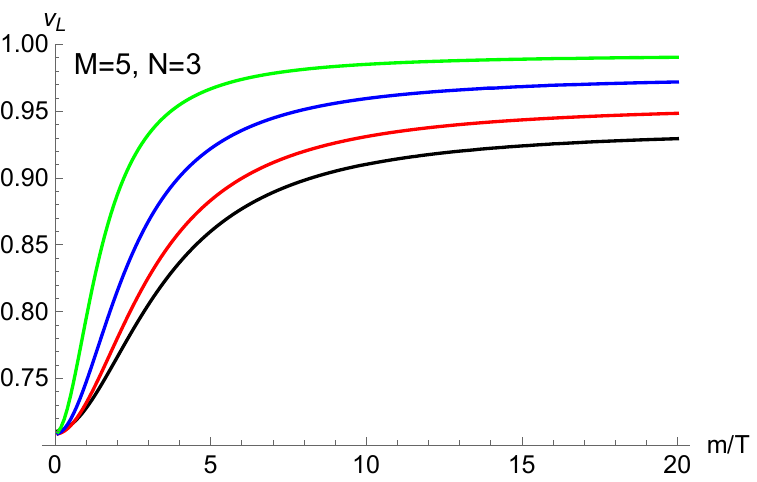}
    \caption{The sound phonon velocity $v_L$ as the function of $m/T$ for $n/m =\{0,1/2,1,2\}$ (black, red, blue, green). When $m/T\to0$, $v_L^2=1/2$ is satisfied. $\mathbf{Left}$: $M=5,N=7$. $\mathbf{Right}$: $M=5,N=3$.}
    \label{fig:v_L}
\end{figure}

\begin{figure}[htbp]
    \centering
    \includegraphics[width=0.48\linewidth]{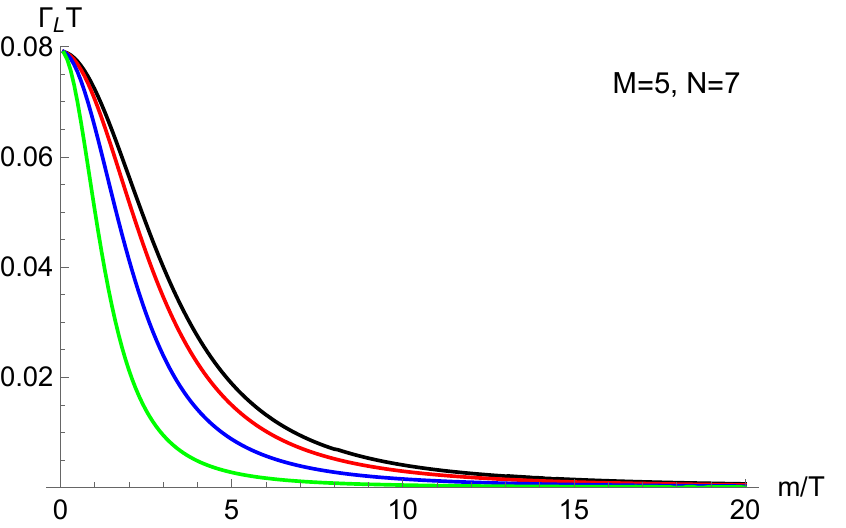}
    \includegraphics[width=0.48\linewidth]{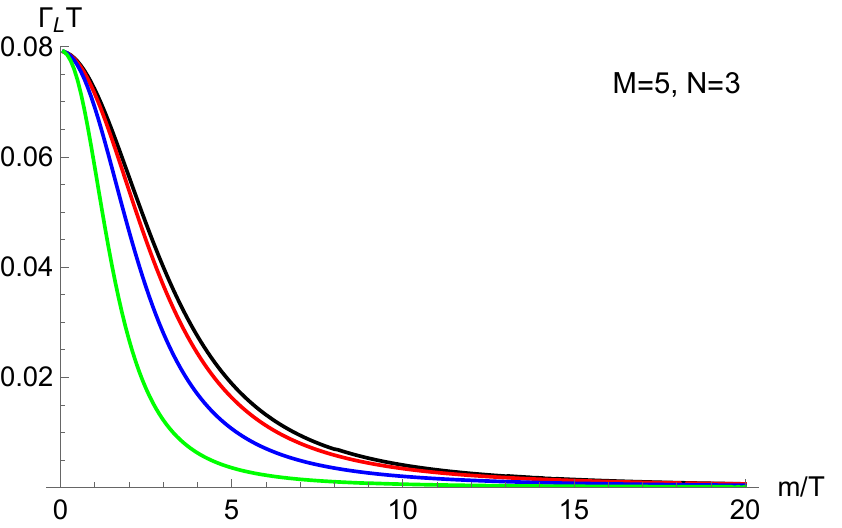}
    \caption{The longitudinal phonon attenuation $\Gamma_L$ as the function of $m/T$ for $n/m =\{0,1/2,1,2\}$ (black, red, blue, green). $\mathbf{Left}$: $M=5,N=7$. $\mathbf{Right}$: $M=5,N=3$.}
    \label{fig:D_L}
\end{figure}

The sound speed and attenuation have been shown in FIG.\ref{fig:v_L} and FIG.\ref{fig:D_L}. One can check that the following relation about the two sound speeds
\begin{equation}
    v_L^2-v_T^2=\frac{1}{2},
\end{equation}
is always ensured due to the conformal symmetry. Then, the speed curve of the longitudinal sound just shifts upwards by $1/2$ with respect to the one of the transverse sound. In addition, the sound attenuation $\Gamma_L$ always gets suppressed in the presence of the $\chi^y$ fields.

The existence of the diffusive mode ``1" in the longitudinal channel has already been revealed in the hydrodynamics of solids long time ago \cite{solidhydro}. In the non-relativistic case, it can be understood as point defect transport process that is not caused by the compressibility of the crystal. And its relativistic version which originates from the coupling between the energy density and the longitudinal Goldstone field at finite temperatures has also been investigated in the relativistic hydrodynamics, holography as well as effective field theories in recent years \cite{Delacretaz:2017zxd,Ammon:2019apj}. In the presence of the $\chi^y$ field, an extra diffusive mode denoted by ``2" appears due to the coupling of the two longitudinal Goldstone fields  corresponding to the bulk fields $\phi^y$ and $\chi^y$. Unlike the transverse diffusion, both of the two diffusive modes obey the proposed diffusion bound.

\begin{figure}[htbp]
    \centering
    \includegraphics[width=0.48\linewidth]{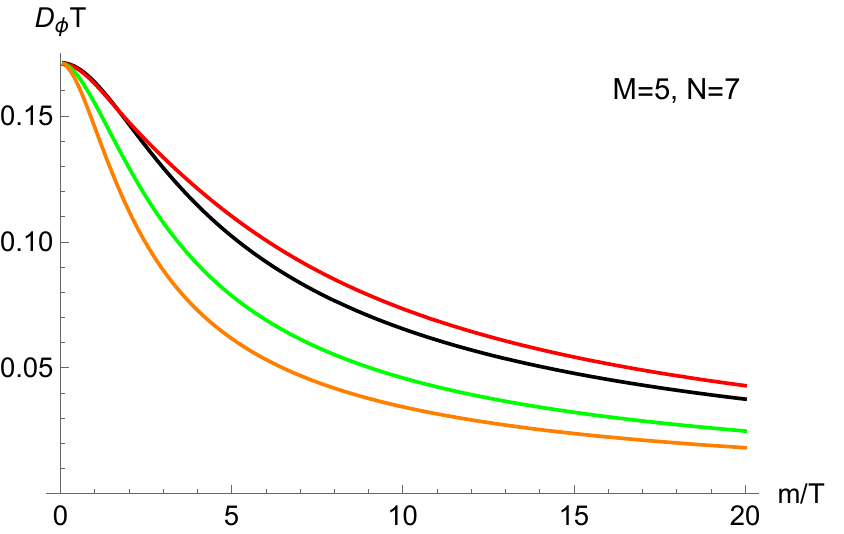}
    \includegraphics[width=0.48\linewidth]{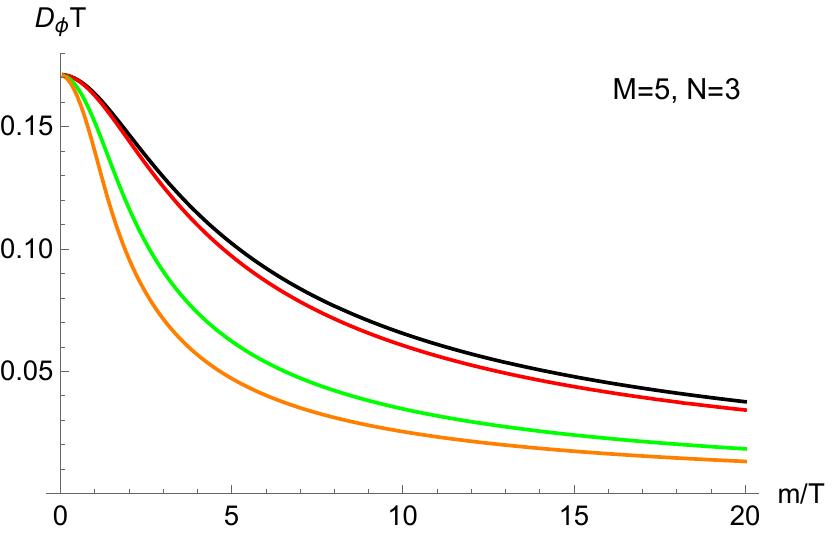}
    \caption{$D_{\phi}\,T$ as the function of $m/T$ for $n/m =\{0,1/2,2,3\}$ (black, red, green, orange). $\mathbf{Left}$: $M=5,N=7$. $\mathbf{Right}$: $M=5,N=3$.}
    \label{fig:Df1}
\end{figure}

\begin{figure}[htbp]
    \centering
    \includegraphics[width=0.48\linewidth]{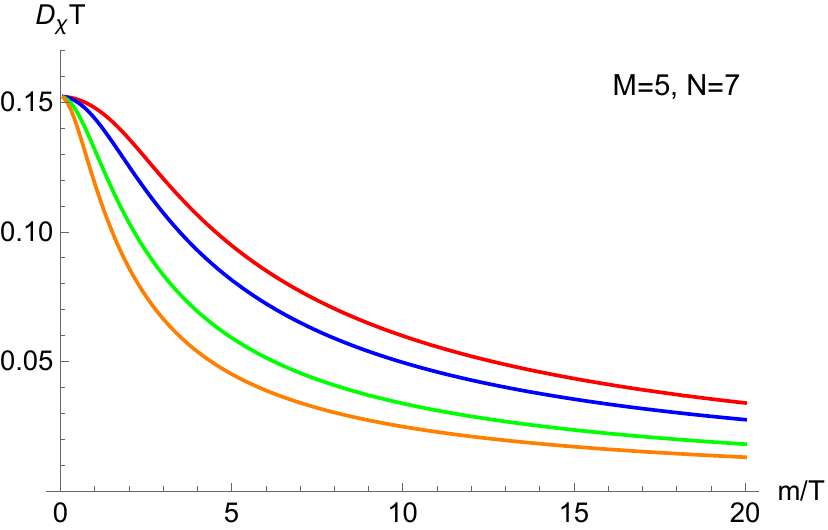}
    \includegraphics[width=0.48\linewidth]{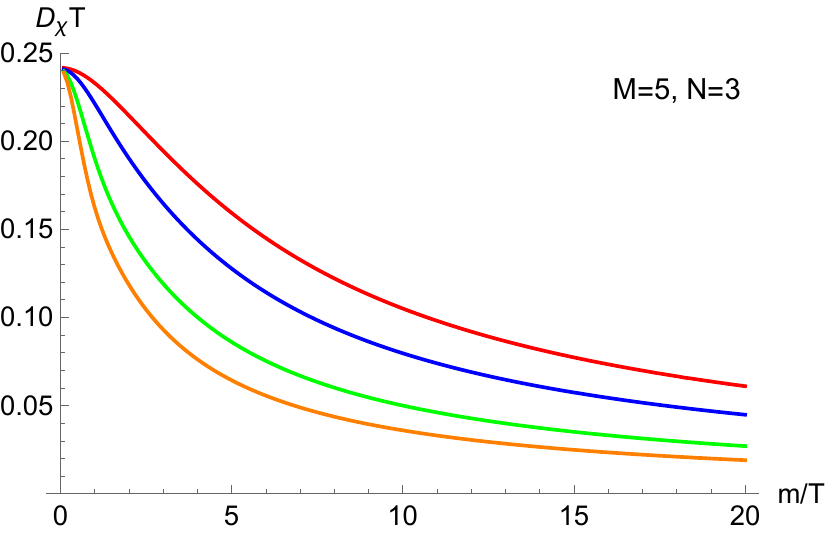}
    \caption{$D_{\chi}\,T$ as the function of $m/T$ for $n/m =\{1/2,1,2,3\}$ (red, blue, green, orange). $\mathbf{Left}$: $M=5,N=7$. $\mathbf{Right}$: $M=5,N=3$.}
    \label{fig:Df2}
\end{figure}

\section{Discussion and outlook}\label{section5}
\begin{figure}[htbp]
    \centering
    \includegraphics[width=0.70\linewidth]{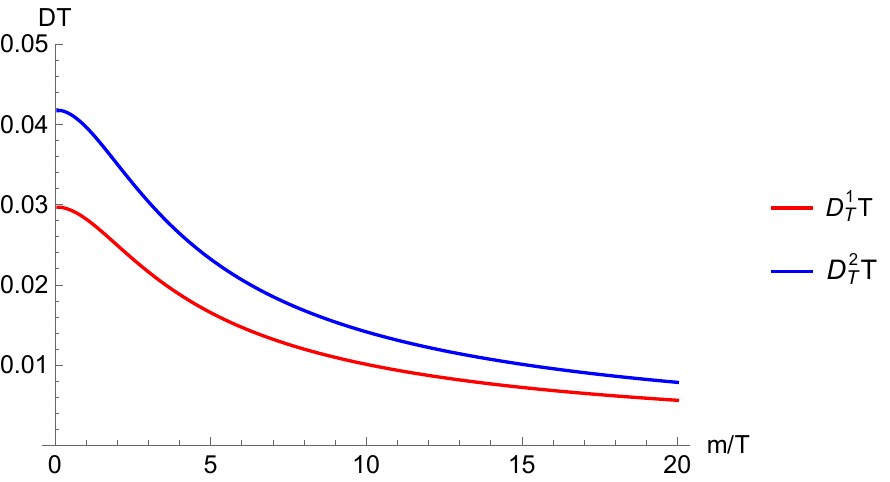}
    \caption{Two transverse diffusive mode emerges in the triple-axion model with $\mathcal{W}=m^2(X^6+Y^5+U^4)$.}
    \label{fig:3-diffusive}
\end{figure}

In this paper, we investigate the shear viscoelasticity as well as the hydrodynamic modes in holographic multiple-axion models where the translations are broken spontaneously. Our results show that adding more sets of axions in the bulk, the shear modulus is enhanced at high temperatures and
shear viscosity is always suppressed.  However, different sets
of axions exhibit competitive relationship in determining the shear modulus at low temperatures. Furthermore, adding more bulk axions will not increase the amount of sound pairs, but just increases diffusive modes on boundary. Although we have so far demonstrated the second point above only in the double-axion model, our conclusion can be easily extended to cases with more axions. As shown in FIG.\ref{fig:3-diffusive}, two diffusive modes emerge in the transverse channel of the triple-axion model. 

Nevertheless, there are still many open questions. For instance, what are the physical natures of the extra diffusive modes in the multiple-axion model? This can be studied by using the method in \cite{Arean:2021tks}. What is the hydrodynamic description of the multiple-axion models? This may be constructed by following  \cite{Armas:2019sbe,Baggioli:2022pyb}. How do the additional axions affect the mechanical stability of the system in the non-linear elastic regime \cite{Baggioli:2023dfj}? We leave these for future work.

\subsection*{Acknowledgments} 
We would like to thank M. Baggioli, S.-Q. Lan, L. Li, H.-T. Sun and X.-J. Wang for helpful discussions. This work is supported by the National Natural Science Foundation of China (NSFC) under Grant No.12275038, the Fundamental Research Funds for the Central Universities No.DUT23BK053, and DUT Innovation and entrepreneurship program for undergraduate students No.20231014110345.

\bibliographystyle{apsrev4-1}
\bibliography{reference}

\newpage
\appendix
\section{Verification of the competitive relationship between $\phi^I$ and $\chi^I$ in determining the shear modulus}
\label{App-G}

\begin{figure}[htbp]
    \centering
    \includegraphics[width=0.7\linewidth]{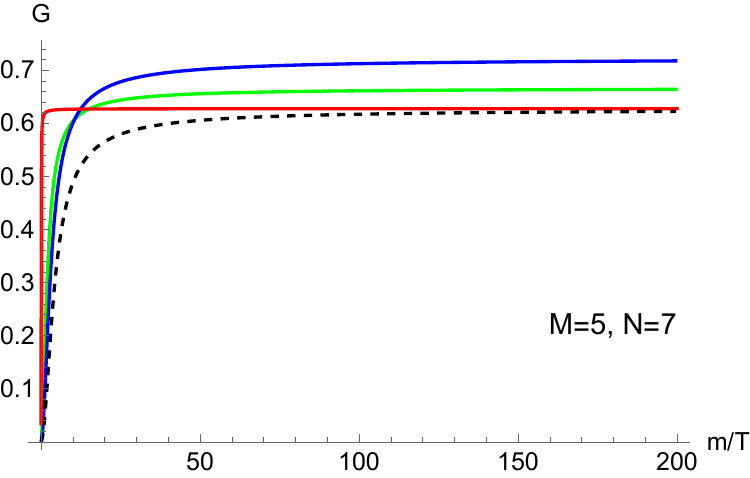}
    \caption{The shear elastic modulus $G$ as the function of $m/T$ for the double-axion model with $\mathcal{W}(X,Y)=m^2X^M+n^2 Y^N$, $n/m$ = $\{1,2,100\}$ (blue, green, red) and the single-axion model with $\mathcal{W}(X)=m^2X^7$ denoted by the dashed line.}
    \label{fig:G-lowT}
\end{figure}

Here, we verify the competitive relationship between the two sets of axions in $G$ at low temperatures which is mentioned in Section \ref{section3}. For $m\gg n$, the contribution from $\chi^I$ fields is just a small correction. Then, the result is expected to be not very different from the single-axion model. Therefore, we only need to consider the case of $m\ll n$. In FIG.\ref{fig:G-lowT}, we plot the $G$-curves for the double-axion model with $\mathcal{W}(X,Y)=m^2X^5+n^2 Y^7$ by varying the value of $n$. It is obvious to see that when $n/m=100$, the result is very close to that in the single-axion model with $\mathcal{W}(X)=m^2 X^7$ at low temperatures which implies that the behavior of $G$ in the double-axion model is solely controlled by the $Y$-term in this regime when $m\ll n$.

\section{Equations of motion of the linearized perturbations}\label{App-a}
\subsection{Transverse sector}
In the radial gauge, $h_{xu}=0$, the EOMs of the transverse modes are given by
\begin{align}
     0=&-2 l^2 u Z _Y  \chi_x' - 2 l^2 u Z _X  \phi_x' - uh_ {tx}'' + 
 2 h_ {tx}' + i k u h_ {xy}' - 2 m^2 u  V_X  \phi_x' - 
 2 n^2 u  W_Y  \chi_x' ,\\
     0=&\ 2 l^2 u^2 h_ {xy}  V _Y + 2 l^2 u^2 h_ {xy}  V _X - 
 2 l^2 h_ {xy}  V - 4 l^2 u^2 h_ {xy}  Z _Y - 
 4 l^2 u^2 h_ {xy}  Z _X + 2 l^2 h_ {xy}  Z+ \nonumber\\
 & 2 i l^2 k u^2 \chi_x  Z _Y + 2 i l^2 k u^2 \phi_x  Z _X + 
 u^2 f'  h_ {xy}' + u^2 f  h_ {xy}'' - 2 u f  h_ {xy}' + 
 i k u^2 h_ {tx}' - 2 i k u h_ {tx} + \nonumber\\&2 i u^2\omega h_ {xy}' - 
 2 m^2 u^2 h_ {xy}   V_X - 2 n^2 u^2 h_ {xy}   W_Y - 
 2 i u\omega h_ {xy} + 2 i k m^2 u^2 \phi_x   V_X + 
 2 i k n^2 u^2 \chi_x   W_Y ,\\
     0=&\ 2 l^2 u^2 f  Z _Y  \chi_x' + 2 l^2 u^2 f  Z _X  \phi_x' + 
 2 l^2 h_ {tx}  V + 2 l^2 u^2 h_ {tx}  Z _Y + 
 2 l^2 u^2 h_ {tx}  Z _X - 2 l^2 h_ {tx}  Z + \nonumber\\&
 2 i l^2 u^2\omega \chi_x  Z _Y + 2 i l^2 u^2\omega \phi_x  Z _X - 
 i k u^2 f  h_ {xy}' + 2 m^2 u^2 f   V_X  \phi_x' + 
 2 n^2 u^2 f   W_Y  \chi_x' - i u^2\omega h_ {tx}' + \nonumber\\&
 k^2 u^2 h_ {tx} + 2 m^2 u^2 h_ {tx}   V_X + 2 n^2 u^2 h_ {tx}   W_Y +
  k u^2\omega h_ {xy} + 2 i m^2 u^2\omega \phi_x   V_X + 
 2 i n^2 u^2\omega \chi_x   W_Y ,\\
     0=&\ l^2 u Z _X  f'  \phi_x' + l^2 u f  Z _X  \phi_x'' - 
 2 l^2 f  Z _X  \phi_x' + 2 l^2 u^2 f  Z_ {XY}  \phi_x' + 
 2 l^2 u^2 f  Z_ {XX}  \phi_x' + l^2 u Z _X  h_ {tx}' - \nonumber\\&
 2 l^2 h_ {tx}  Z _X + 2 l^2 u^2 h_ {tx}  Z_ {XY} + 
 2 l^2 u^2 h_ {tx}  Z_ {XX} - i l^2 k u h_ {xy}  Z _X - 
 l^2 k^2 u \phi_x  Z _X + 2 i l^2 u\omega Z _X  \phi_x' - \nonumber\\&
 2 i l^2\omega \phi_x  Z _X + 2 i l^2 u^2\omega \phi_x  Z_ {XY} + 
 2 i l^2 u^2\omega \phi_x  Z_ {XX} + m^2 u f'   V_X  \phi_x' + 
 2 m^2 u^2 f   V_X'  \phi_x' + \nonumber\\&m^2 u f   V_X  \phi_x'' - 
 2 m^2 f   V_X  \phi_x' + m^2 u h_ {tx}'   V_X + 
 2 m^2 u^2 h_ {tx}   V_X' - 2 m^2 h_ {tx}   V_X - 
 i k m^2 u h_ {xy}   V_X - \nonumber\\&k^2 m^2 u \phi_x   V_X + 
 2 i m^2 u^2\omega \phi_x   V_X' + 2 i m^2 u\omega  V_X  \phi_x' - 
 2 i m^2\omega \phi_x   V_X ,\\
      0=&l^2 u Z _Y  f'  \chi_x' + l^2 u f  Z _Y  \chi_x'' - 
 2 l^2 f  Z _Y  \chi_x' + 2 l^2 u^2 f  Z_ {YY}  \chi_x' + 
 2 l^2 u^2 f  Z_ {XY}  \chi_x' + l^2 u Z _Y  h_ {tx}' - \nonumber\\&
 2 l^2 h_ {tx}  Z _Y + 2 l^2 u^2 h_ {tx}  Z_ {YY} + 
 2 l^2 u^2 h_ {tx}  Z_ {XY} - i l^2 k u h_ {xy}  Z _Y - 
 l^2 k^2 u \chi_x  Z _Y + 2 i l^2 u\omega Z _Y  \chi_x' - \nonumber\\&
 2 i l^2\omega \chi_x  Z _Y + 2 i l^2 u^2\omega \chi_x  Z_ {YY} + 
 2 i l^2 u^2\omega \chi_x  Z_ {XY} + n^2 u f'   W_Y  \chi_x' + 
 2 n^2 u^2 f   W_Y'  \chi_x' +\nonumber\\& n^2 u f   W_Y  \chi_x'' - 
 2 n^2 f   W_Y  \chi_x' + n^2 u h_ {tx}'   W_Y + 
 2 n^2 u^2 h_ {tx}   W_Y' - 2 n^2 h_ {tx}   W_Y - 
 i k n^2 u h_ {xy}   W_Y - \nonumber\\&k^2 n^2 u \chi_x   W_Y + 
 2 i n^2 u^2\omega \chi_x   W_Y' + 2 i n^2 u\omega  W_Y  \chi_x' - 
 2 i n^2\omega \chi_x   W_Y ,
\end{align}
where the prime denotes the partial derivative with respect to the radial coordinate $u$ and the subscripts `$X$' and `$Y$' represent $\partial/\partial X$ and $\partial/\partial Y$.
\subsection{Longitudinal sector}
For the longitudinal sector, we introduce the symmetric combination $h_{xs}=(h_{xx}+h_{yy})/2$ and the anti-symmetric combination $h_{xa}=(h_{xx}-h_{yy})/2$ to simplify the equations. In the the radial gauge, $h_{uu}=h_{tu}=h_{yu}=0$, we have that
\begin{align}
    0=&\ h_{xs}^{\prime\prime},\\
    0=&\ 2 l^2 u Z _Y  \chi_y' + 2 l^2 u Z _X  \phi_y' + u h_ {ty}'' - 
 2 h_ {ty}' + i k u h_ {xa}' + i k u h_ {xs}' + 
 2 m^2 u  V_X  \phi_y' + 2 n^2 u  W_Y  \chi_y' ,\\
    0=&- l^2 u f  Z  h_ {xs}' + 2 l^2 u^2 f  h_ {xs}  Z _Y + 
 2 l^2 u^2 f  h_ {xs}  Z _X - 2 i l^2 k u^2 f  \chi_y  Z _Y - 
 2 i l^2 k u^2 f  \phi_y  Z _X -\nonumber\\& i l^2 k u h_ {ty}  Z - 
  i l^2 u\omega h_ {xs}  Z - 2 u f  h_ {tt}' + 6 f  h_ {tt} + 
 i k u f  h_ {ty} +  k^2 u^2 f  h_ {xa} -  m^2 u f  V  h_ {xs}' - \nonumber\\&
  n^2 u f  W  h_ {xs}' +3 u f  h_ {xs}' +  u f ^2 h_ {xs}' + 
  k^2 u^2 f  h_ {xs} + 2 m^2 u^2 f  h_ {xs}   V_X + 
2 n^2 u^2 f  h_ {xs}   W_Y +\nonumber\\& i u\omega f  h_ {xs} - 
2 i k m^2 u^2 f  \phi_y   V_X - 2 i k n^2 u^2 f  \chi_y   W_Y - 
  k^2 u^2 h_ {tt} + 2 i u\omega h_ {tt} -  i k m^2 u h_ {ty}  V - \nonumber\\&
  i k n^2 u h_ {ty}  W - 2 k u^2\omega h_ {ty} + 3 i k u h_ {ty} - 
  i m^2 u\omega h_ {xs}  V -  i n^2 u\omega h_ {xs}  W - 
 2 u^2\omega^2 h_ {xs} + 3 i u\omega h_ {xs} ,\\
    0=&\ 2 l^2 u^2 f  Z _Y  \chi_y' + 2 l^2 u^2 f  Z _X  \phi_y' + 
 2 l^2 u^2 h_ {ty}  Z _Y + 2 l^2 u^2 h_ {ty}  Z _X + 
 2 i l^2 u^2\omega \chi_y  Z _Y + 2 i l^2 u^2\omega \phi_y  Z _X \nonumber\\&+ 
 i k u^2 f  h_ {xa}' + i k u^2 f  h_ {xs}' + 
 2 m^2 u^2 f   V_X  \phi_y' + 2 n^2 u^2 f   W_Y  \chi_y' - 
 i k u^2 h_ {tt}' + 2 i k u h_ {tt} - i u^2\omega h_ {ty}' + \nonumber\\&
 2 m^2 u^2 h_ {ty}   V_X + 2 n^2 u^2 h_ {ty}   W_Y - 
 k u^2\omega h_ {xa} - k u^2\omega h_ {xs} + 
 2 i m^2 u^2\omega \phi_y   V_X + 2 i n^2 u^2\omega \chi_y   W_Y ,\\
    0=&-2 l^2 u^2 h_ {xa}   Z_ {Y} - 2 l^2 u^2 h_ {xa}   Z_ {X} - 
 2 l^2 u^4 h_ {xs}   Z_ {YY} - 4 l^2 u^4 h_ {xs}   Z_ {XY} - 
 2 l^2 u^4 h_ {xs}   Z_ {XX} - \nonumber\\&2 i l^2 k u^2   \chi_y   Z_ {Y} - 
 2 i l^2 k u^2   \phi_y   Z_ {X} + 2 i l^2 k u^4   \chi_y   Z_ {YY} + 
 2 i l^2 k u^4   \chi_y   Z_ {XY} + 
 2 i l^2 k u^4   \phi_y   Z_ {XY} + \nonumber\\&2 i l^2 k u^4   \phi_y   Z_ {XX} +
  u^2 f'   h_ {xa}'  - u^2 f'   h_ {xs}'  + u^2 f   h_ {xa}''  - 
 2 u f   h_ {xa}'  - u^2 f   h_ {xs}''  + 2 u f   h_ {xs}'  + 
 u^2 h_ {tt}''- \nonumber\\&  4 u h_ {tt}'  + 6 h_ {tt}  - 2 i k u^2 h_ {ty}'  + 
 4 i k u h_ {ty}  + 2 i u^2\omega h_ {xa}'  - 
 2 m^2 u^2 h_ {xa}    V_ {X} - 2 n^2 u^2 h_ {xa}    W_ {Y} - \nonumber\\&
 2 i u\omega h_ {xa}  - 2 i u^2\omega h_ {xs}'  - 
 2 m^2 u^4 h_ {xs}    V_ {X}' - 2 n^2 u^4 h_ {xs}    W_ {Y}' + 
 2 i u\omega h_ {xs}  - 2 i k m^2 u^2   \phi_y    V_ {X} + \nonumber\\&
 2 i k m^2 u^4   \phi_y    V_ {X} ' - 
 2 i k n^2 u^2   \chi_y    W_ {Y} + 
 2 i k n^2 u^4   \chi_y    W_ {Y} ' ,
 \end{align}
 \begin{align}
    0=&\ 2 l^2 u^2 h_ {xa}   Z_ {Y} + 2 l^2 u^2 h_ {xa}   Z_ {X} - 
 2 l^2 u^4 h_ {xs}   Z_ {YY} - 4 l^2 u^4 h_ {xs}   Z_ {XY} - 
 2 l^2 u^4 h_ {xs}   Z_ {XX} + \nonumber\\&2 i l^2 k u^2   \chi_y   Z_ {Y} + 
 2 i l^2 k u^2   \phi_y   Z_ {X} + 2 i l^2 k u^4   \chi_y   Z_ {YY} + 
 2 i l^2 k u^4   \chi_y   Z_ {XY} + 
 2 i l^2 k u^4   \phi_y   Z_ {XY} + \nonumber\\&2 i l^2 k u^4   \phi_y   Z_ {XX} -
  u^2 f'   h_ {xa}'  - u^2 f'   h_ {xs}'  - u^2 f   h_ {xa}''  + 
 2 u f   h_ {xa}'  - u^2 f   h_ {xs}''  + 2 u f   h_ {xs}'  + 
 u^2 h_ {tt}''  - \nonumber\\&4 u h_ {tt}'  + 6 h_ {tt}  - 
 2 i u^2\omega h_ {xa}'  + 2 m^2 u^2 h_ {xa}    V_ {X} + 
 2 n^2 u^2 h_ {xa}    W_ {Y} + 2 i u\omega h_ {xa}  - 
 2 i u^2\omega h_ {xs}'  - \nonumber\\&2 m^2 u^4 h_ {xs}    V_ {X}' - 
 2 n^2 u^4 h_ {xs}    W_ {Y}' + 2 i u\omega h_ {xs}  + 
 2 i k m^2 u^2   \phi_y    V_ {X} + 
 2 i k m^2 u^4   \phi_y    V_ {X} ' + 
 2 i k n^2 u^2   \chi_y    W_ {Y} \nonumber\\&+ 
 2 i k n^2 u^4   \chi_y    W_ {Y} ' ,\\
    0=&\ l^2 u Z  h_ {xs}'  - 2 l^2 u^2 h_ {xs}   Z_ {Y} - 
 2 l^2 u^2 h_ {xs}   Z_ {X} + 2 i l^2 k u^2   \chi_y   Z_ {Y} + 
 2 i l^2 k u^2   \phi_y   Z_ {X} - u f   h_ {xs}'  + 2 u h_ {tt}'  - \nonumber\\&
 6 h_ {tt}  + i k u^2 h_ {ty}'  - 4 i k u h_ {ty}  - 
 k^2 u^2 h_ {xa}  + m^2 u V  h_ {xs}'  + n^2 u W  h_ {xs}'  + 
 2 i u^2\omega h_ {xs}'  - 3 u h_ {xs}'  - \nonumber\\&k^2 u^2 h_ {xs}  - 
 2 m^2 u^2 h_ {xs}    V_ {X} - 2 n^2 u^2 h_ {xs}    W_ {Y} - 
 4 i u\omega h_ {xs}  + 2 i k m^2 u^2   \phi_y    V_ {X} + 
 2 i k n^2 u^2   \chi_y    W_ {Y} ,\\
    0=&\ l^2 u Z_ {X}  f'     \phi_y'  + l^2 u f   Z_ {X}    \phi_y''  + 
 2 l^2 u^2 f   Z_ {XY}    \phi_y'  + 
 2 l^2 u^2 f   Z_ {XX}    \phi_y'  - 2 l^2 f   Z_ {X}    \phi_y'  + 
 l^2 u Z_ {X}  h_ {ty}'  +\nonumber\\& 2 l^2 u^2 h_ {ty}   Z_ {XY} + 
 2 l^2 u^2 h_ {ty}   Z_ {XX} - 2 l^2 h_ {ty}   Z_ {X} + 
 i l^2 k u h_ {xa}   Z_ {X} - i l^2 k u^3 h_ {xs}   Z_ {XY} - 
 i l^2 k u^3 h_ {xs}   Z_ {XX} \nonumber\\&-l^2 k^2 u   \phi_y   Z_ {X} - 
 l^2 k^2 u^3   \chi_y   Z_ {XY} - l^2 k^2 u^3   \phi_y   Z_ {XX} + 
 2 i l^2 u\omega Z_ {X}    \phi_y'  + 
 2 i l^2 u^2\omega   \phi_y   Z_ {XY} + \nonumber\\&
 2 i l^2 u^2\omega   \phi_y   Z_ {XX} - 
 2 i l^2\omega   \phi_y   Z_ {X} + m^2 u f'    V_ {X}    \phi_y'  + 
 2 m^2 u^2 f    V_ {X} '    \phi_y'  + 
 m^2 u f    V_ {X}    \phi_y''  - 2 m^2 f    V_ {X}    \phi_y'  + \nonumber\\&
 m^2 u h_ {ty}'    V_ {X} + 2 m^2 u^2 h_ {ty}    V_ {X}' - 
 2 m^2 h_ {ty}    V_ {X} + i k m^2 u h_ {xa}    V_ {X} - 
 i k m^2 u^3 h_ {xs}    V_ {X}' - k^2 m^2 u   \phi_y    V_ {X} - \nonumber\\&
 k^2 m^2 u^3   \phi_y    V_ {X} ' + 
 2 i m^2 u^2\omega   \phi_y    V_ {X} ' + 
 2 i m^2 u\omega  V_ {X}    \phi_y'  - 
 2 i m^2\omega   \phi_y    V_ {X} ,\\
    0=&\ l^2 u Z_ {Y}  f'     \chi_y'  + l^2 u f   Z_ {Y}    \chi_y''  + 
 2 l^2 u^2 f   Z_ {YY}    \chi_y'  + 
 2 l^2 u^2 f   Z_ {XY}    \chi_y'  - 2 l^2 f   Z_ {Y}    \chi_y'  + 
 l^2 u Z_ {Y}  h_ {ty}'  + \nonumber\\&2 l^2 u^2 h_ {ty}   Z_ {YY} + 
 2 l^2 u^2 h_ {ty}   Z_ {XY} - 2 l^2 h_ {ty}   Z_ {Y} + 
 i l^2 k u h_ {xa}   Z_ {Y} - i l^2 k u^3 h_ {xs}   Z_ {YY} - 
 i l^2 k u^3 h_ {xs}   Z_ {XY}\nonumber\\& - l^2 k^2 u   \chi_y   Z_ {Y} - 
 l^2 k^2 u^3   \chi_y   Z_ {YY} - l^2 k^2 u^3   \phi_y   Z_ {XY} + 
 2 i l^2 u\omega Z_ {Y}    \chi_y'  + 
 2 i l^2 u^2\omega   \chi_y   Z_ {YY} + \nonumber\\&
 2 i l^2 u^2\omega   \chi_y   Z_ {XY} - 
 2 i l^2\omega   \chi_y   Z_ {Y} + n^2 u f'    W_ {Y}    \chi_y'  + 
 2 n^2 u^2 f    W_ {Y} '    \chi_y'  + 
 n^2 u f    W_ {Y}    \chi_y''  - 2 n^2 f    W_ {Y}    \chi_y'  + \nonumber\\&
 n^2 u h_ {ty}'    W_ {Y} + 2 n^2 u^2 h_ {ty}    W_ {Y}' - 
 2 n^2 h_ {ty}    W_ {Y} + i k n^2 u h_ {xa}    W_ {Y} - 
 i k n^2 u^3 h_ {xs}    W_ {Y}' - k^2 n^2 u   \chi_y    W_ {Y} - \nonumber\\&
 k^2 n^2 u^3   \chi_y    W_ {Y} ' + 
 2 i n^2 u^2\omega   \chi_y    W_ {Y} ' + 
 2 i n^2 u\omega  W_ {Y}    \chi_y'  - 
 2 i n^2\omega   \chi_y    W_ {Y} .
\end{align}




\section{Some results for the double-axion model with interaction terms}\label{App-d}
Now, let us turn on the interaction term as follows
\begin{equation}
 \mathcal{W}(X,Y)=m^2X^M+n^2Y^N+l^2X^P\,Y^Q 
\end{equation}
with a non-zero coupling constant $l$ and check how it changes the shear viscoelasticity and the shear hydrodynamic modes by varying its value. In all the cases, we will fix $n/m=1,\,2$ and will show that the interaction term does not change the whole story but just modify the behavior of these quantities quantitatively. 

\begin{figure}[htbp]
    \centering
    \includegraphics[width=0.48\linewidth]{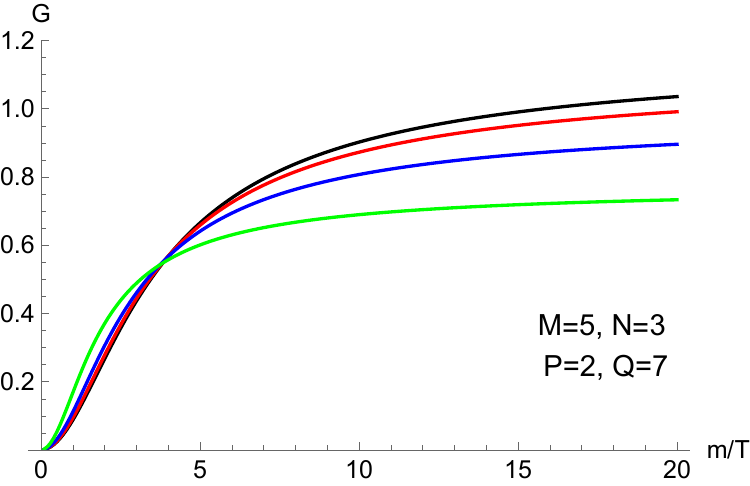}
    \includegraphics[width=0.48\linewidth]{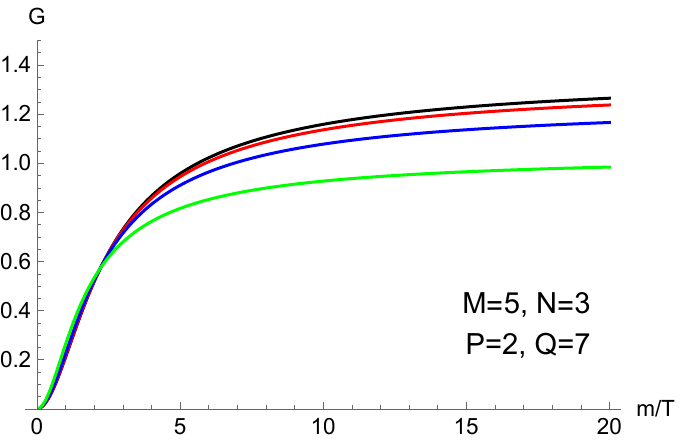}
    \caption{The shear elastic modulus $G$ as the function of $m/T$ for $l/m =\{0,1/2,1,2\}$ (from black to green). We fix $M=5,N=3$ and $P=2,Q=7$. $\textbf{Left}$: $n/m=1$. $\textbf{Right}$: $n/m=2$.}
    \label{fig:G_l}
\end{figure}

\begin{figure}[htbp]
    \centering
    \includegraphics[width=0.48\linewidth]{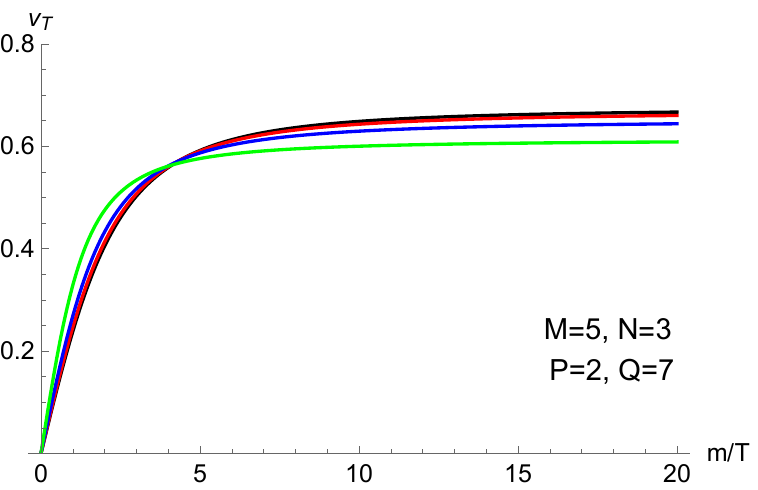}
    \includegraphics[width=0.48\linewidth]{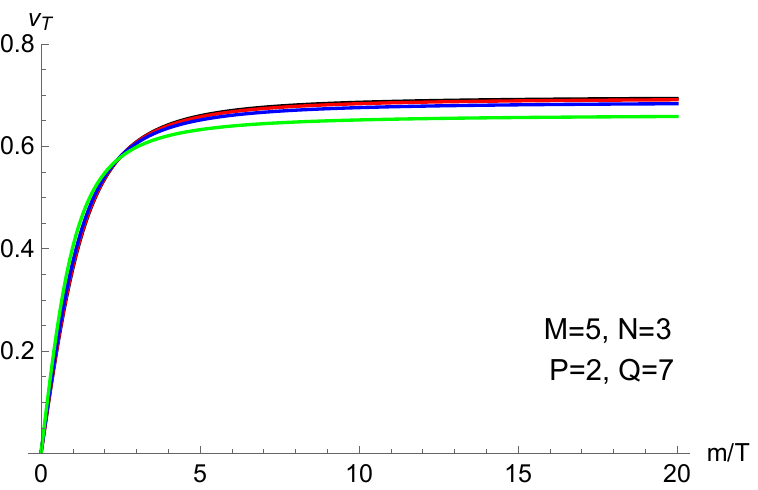}
    \caption{The transverse sound speed $v_T$ as the function of $m/T$ for $l/m =\{0,1/2,1,2\}$ (from black to green). We fix $M=5,N=3$ and $P=2,Q=7$. $\textbf{Left}$: $n/m=1$. $\textbf{Right}$: $n/m=2$.}
    \label{fig:vT_l}
\end{figure}

\begin{figure}[htbp]
    \centering
    \includegraphics[width=0.48\linewidth]{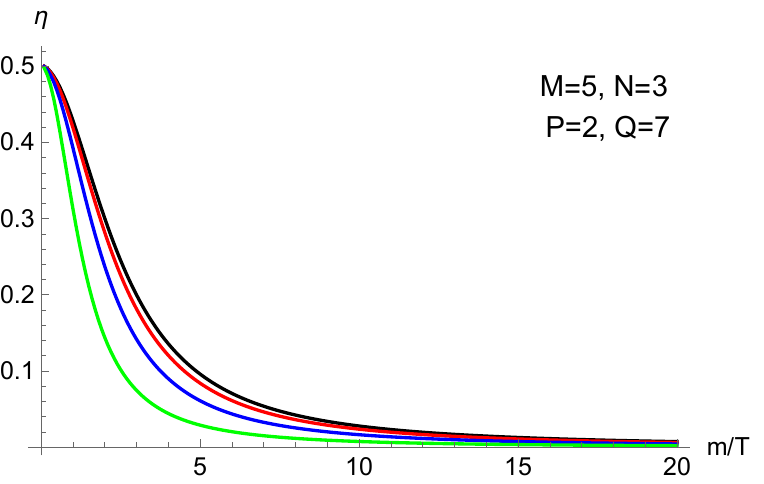}
    \includegraphics[width=0.48\linewidth]{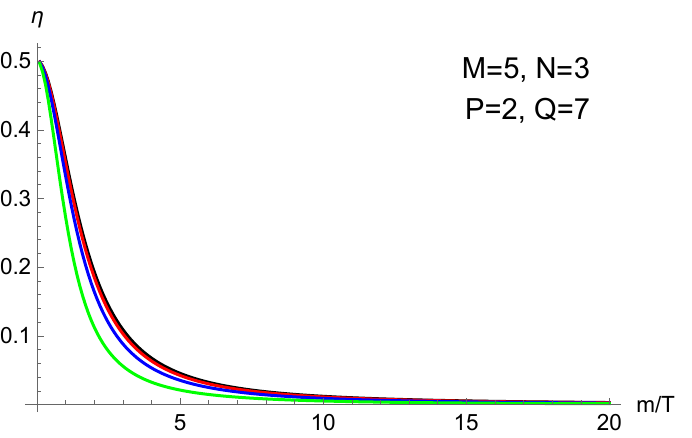}
    \caption{The shear viscosity $\eta$ as the function of $m/T$ for $l/m =\{0,1/2,1,2\}$ (from black to green). We fix $M=5,N=3$ and $P=2,Q=7$. $\textbf{Left}$: $n/m=1$. $\textbf{Right}$: $n/m=2$.}
    \label{fig:eta_l}
\end{figure}

\begin{figure}[htbp]
    \centering
    \includegraphics[width=0.48\linewidth]{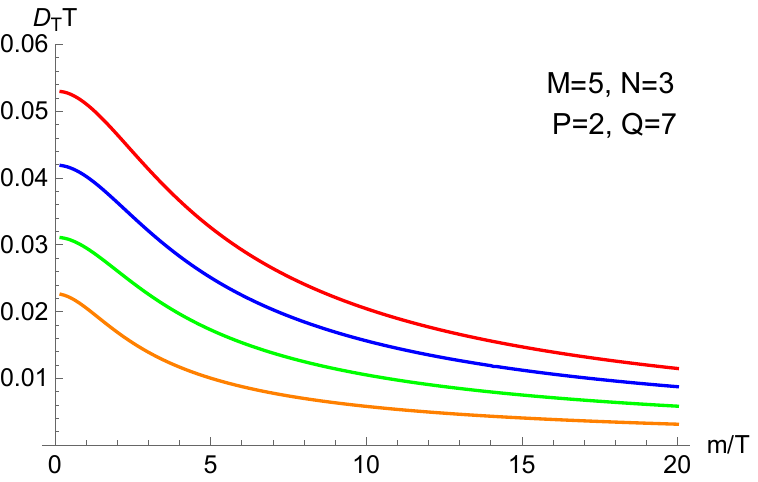}
    \includegraphics[width=0.48\linewidth]{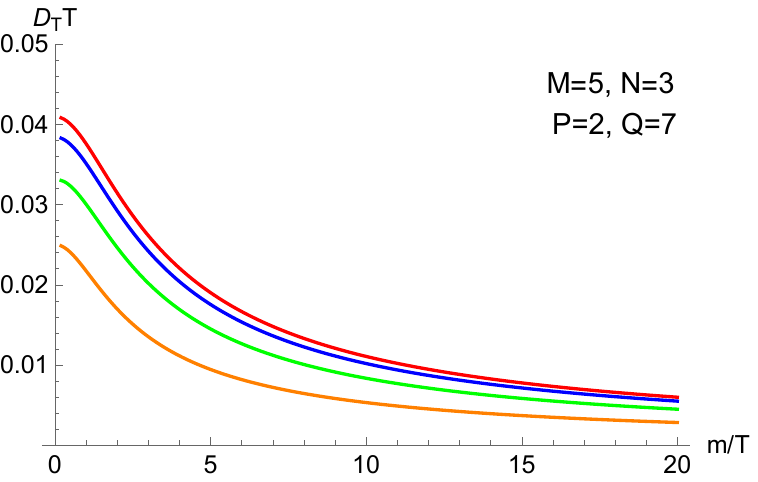}
    \caption{The transverse diffusivity $D_T$ as the function of $m/T$ for $l/m =\{1/2,1,2,3\}$ (from red to orange). We fix $M=5,N=3$ and $P=2,Q=7$. $\textbf{Left}$: $n/m=1$. $\textbf{Right}$: $n/m=2$.}
    \label{fig:DT_l2}
\end{figure}

More precisely, in FIG.\ref{fig:G_l}, we plot the shear elastic modulus $G$ as the function of $m/T$ for different values of $l/m$. In the presence of the interaction between $\phi^I$ and $\chi^I$, the shear modulus $G$ still increases monotonously as the increase of $m/T$. Furthermore, $G$ is enhanced at high temperatures (i.e., when $m/T$ is small), while it is suppressed at low temperatures due to the interaction. Since the shear modulus determines the propagation of the transverse modes, the sound speed $v_T$, as is shown in FIG.\ref{fig:vT_l}, behaves similarly as the shear modulus. Finally, we check how dissipative coefficients, including the shear viscosity, the sound attenuation as well as the diffusion constant, are influenced by the interaction term. In all cases, we find that they all decrease as the increase of $m/T$ and are always suppressed by the interaction. The results have been plotted in FIG.\ref{fig:eta_l}-FIG.\ref{fig:DT_l2}.

\begin{figure}[htbp]
    \centering
    \includegraphics[width=0.48\linewidth]{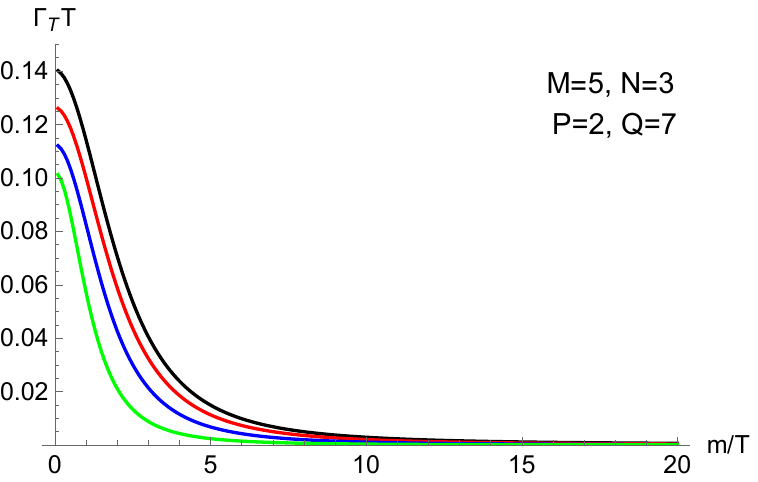}
    \includegraphics[width=0.48\linewidth]{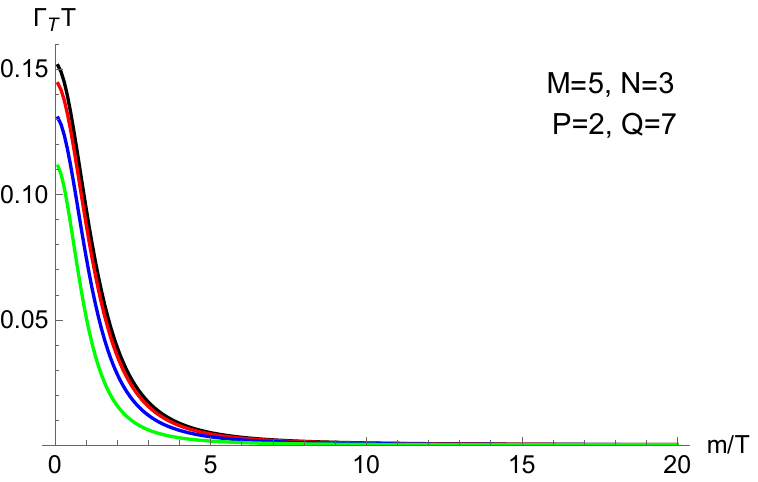}
    \caption{The transverse sound attenuation $\Gamma_T$ as the function of $m/T$ for $l/m =\{0,1/2,1,2\}$ (from black to green). We fix $M=5,N=3$ and $P=2,Q=7$. $\textbf{Left}$: $n/m=1$. $\textbf{Right}$: $n/m=2$.}
    \label{fig:DT_l1}
\end{figure}

\section{A check about the diffusion bound}\label{App-e}

In this section, we verify the bounds for diffusive modes in both of transverse and longitudinal channels. According to \cite{Baggioli:2020ljz}, the butterfly velocity $v_B$ is expressed as
\begin{equation}\label{butterfly}
    v_B^2=\frac{\pi T}{u_h}.
\end{equation}

\begin{figure}[htbp]
    \centering
    \includegraphics[width=0.45\linewidth]{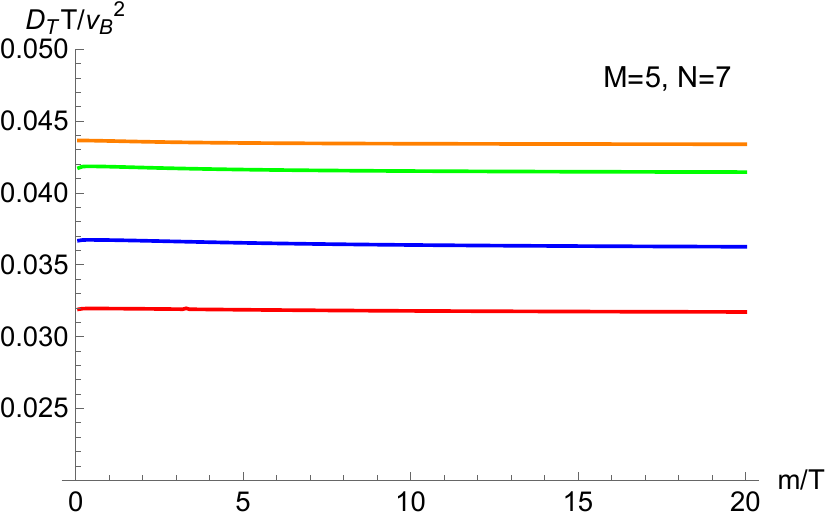}
\includegraphics[width=0.45\linewidth]{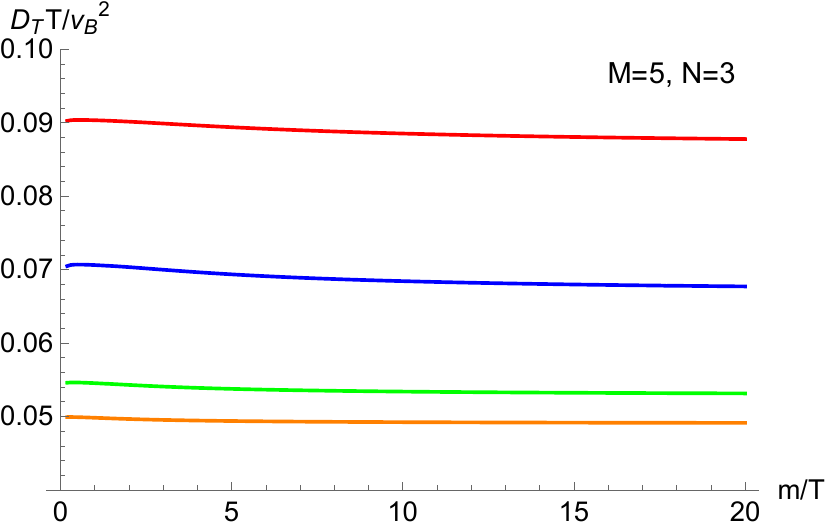}
    \caption{The ratio $D_T T/v_B^2$ as the function of $m/T$ for $n/m =\{1/2,1,2,3\}$ (red, blue, green, orange). $\textbf{Left}$: $M=5,N=7$.  $\textbf{Right}$: $M=5,N=3$.}
    \label{bound-T}
\end{figure}

Then, combining (\ref{butterfly}) and the results for the diffusion constants in the main text, the ratio $D_T T/v_B^2$ as the function of $m/T$ is plotted in FIG.\ref{bound-T}, exhibiting a gradual decrease as the increase of $m/T$. Furthermore, this ratio sensitively depends on the value of $n/m$. For $M<N$, the ratio is enhanced as the increase of $n/m$, while it is suppressed as the increase of $n/m$ for $M>N$. In both cases, we find that $D_T T/v_B^2\ll1$ which implies that the proposed diffusion bound is strongly violated.

\begin{figure}[htbp]
    \centering
    \includegraphics[width=0.45\linewidth]{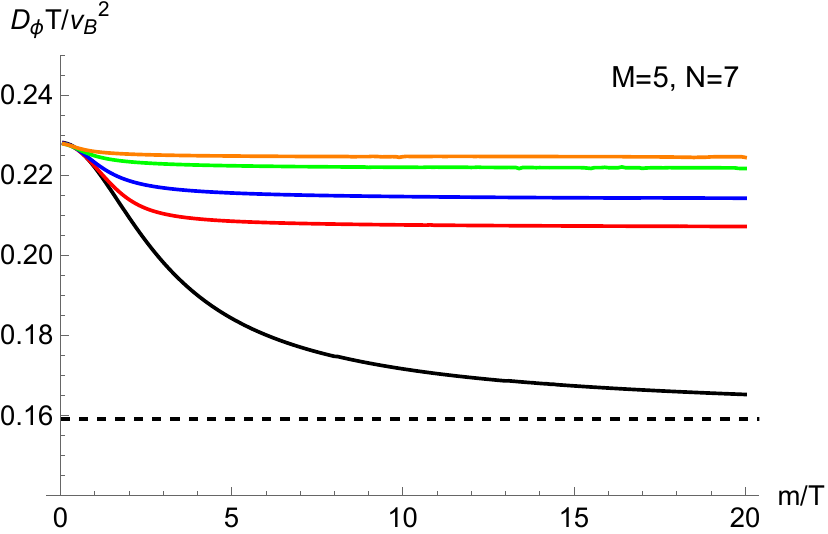}
    \includegraphics[width=0.45\linewidth]{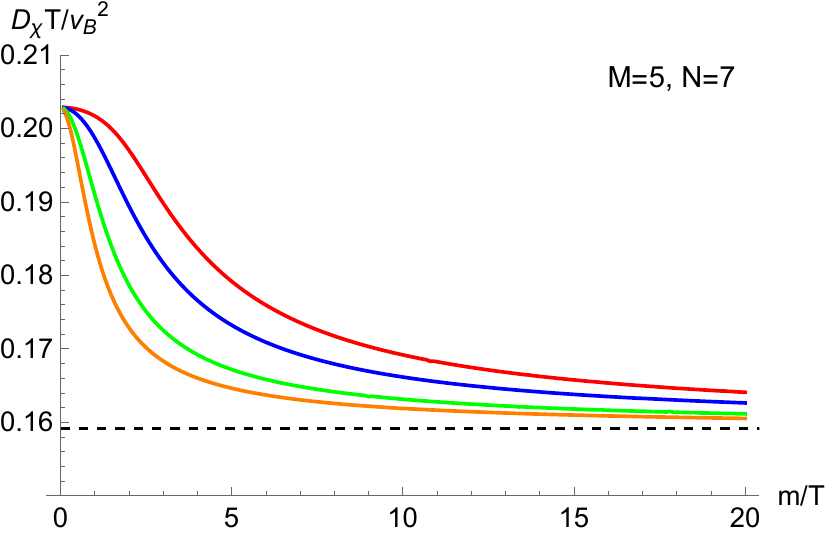}
    \caption{The ratios $D_\phi T/v_B^2$ and $D_\chi T/v_B^2$  as functions of $m/T$ for the double-axion model with $M=5$ and $N=7$. Here, we set $n/m =\{0,1/2,1,2,3\}$ (black, red, blue, green, orange). Dashed line: $\text{AdS}_2$ bound.}
    \label{bound-T1}
\end{figure}

\begin{figure}[htbp]
    \centering
     \includegraphics[width=0.45\linewidth]{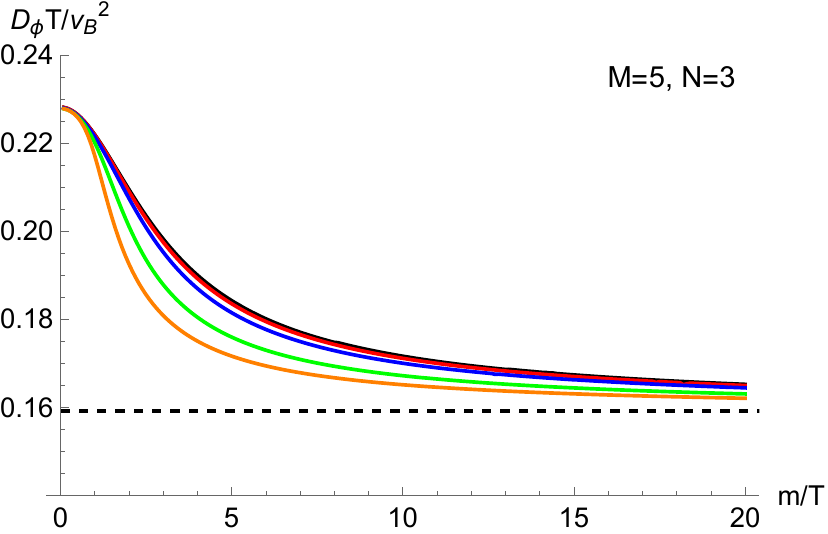}
    \includegraphics[width=0.45\linewidth]{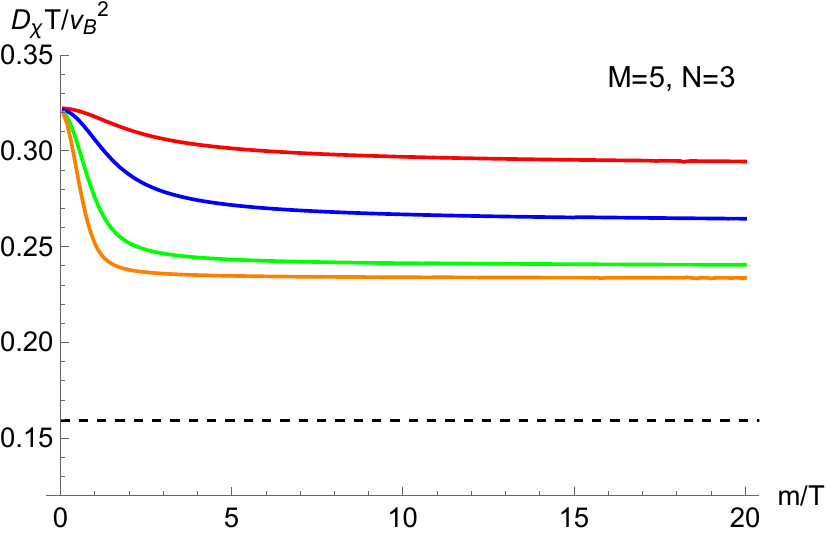}
    \caption{The ratios $D_\phi T/v_B^2$ and $D_\chi T/v_B^2$  as functions of $m/T$ for the double-axion model with $M=5$ and $N=3$. Here, we set $n/m =\{0,1/2,1,2,3\}$ (black,red, blue, green, orange).  Dashed line: $\text{AdS}_2$ bound.}
    \label{bound-T2}
\end{figure}

Finally, the two longitudinal diffusive modes behave in the opposite way as varying the value of $n/m$. The results have been shown in 
FIG.\ref{bound-T1} and FIG.\ref{bound-T2}. For $M<N$, the value of $D_\phi T/v_B^2$ increases as the increase of $n/m$. While $D_\chi T/v_B^2$ always decreases monotonously, approaching the $\text{AdS}_2$ bound in the low temperature limit for all the values of $n/m$. Since this model has a symmetry of $X,M,m \leftrightarrow Y,N,n$, it is found that the results are also symmetric for $M>N$ if the exchange $D_\phi\leftrightarrow D_\chi$ is taken.
\end{document}